\documentclass{blois}

\def\be{\begin{equation}}
\def\ee{\end{equation}}
\def\bea{\begin{eqnarray}}
\def\eea{\end{eqnarray}}

\newcommand{\Photo}{}

\begin{document}
\vspace*{4cm}
\title{THE ALICE EXPERIMENT UPGRADES}

\author{A. Ferretti }

\address{Dipartimento di Fisica dell'Università di Torino, Via Giuria 1,\\
10125 Torino, Italy}

\maketitle\abstracts{
The ALICE experiment profited of the Long Shutdown during 2019-2021 in order to expand its physics capabilities and fully profit from the increased LHC luminosity in Run 3. The Inner Tracking System has been replaced with a new silicon tracker based on MAPS technology, and a new tracking device has been added in front of the Muon Spectrometer to improve its vertexing capabilities. The wire chambers for TPC readout have been replaced with new GEM detectors which will minimize ion backflow and allow for continuous data taking: moreover, a new detector array dedicated to fast triggering has been installed. On the software side, a new first pass reconstruction was added in order to handle and reduce the data flow and storage. These upgrades will be presented together with an outlook of the future ALICE upgrades in view of the LHC Run 4, which will include the replacement of the ITS inner tracking layers with upgraded silicon devices and a high-granularity electromagnetic and hadronic calorimeter in the forward direction (FOCAL)}

\section{The ALICE upgrades for Run 3 and Run 4}

ALICE~\cite{al} is one of the main four experiments at the CERN LHC and is mainly devoted to heavy-ion physics studies to perform a detailed characterization of the strongly interacting quark-gluon plasma (QGP). It is designed to carry out comprehensive studies of hadrons, electrons, muons, heavy flavors, photons and jets produced in heavy-ion collisions on a large kinematic range.

After the successful Run~1 and Run~2 data taking, ALICE profited of the 2019-2021 Long Shutdown to perform a comprehensive upgrade\cite{up} of its apparatus which will allow to improve statistics and resolution of measurements of observables such as heavy-flavour mesons and baryons (down to very low transverse momentum), charmonium states, dileptons from QGP radiation and low-mass vector mesons, light nuclei and hypernuclei.

Since no dedicated trigger to select these events is feasible, ALICE collaboration has decided to pursue a minimum-bias data-taking strategy and to increase the event readout rate for Pb-Pb interactions from 1~kHz to 50~kHz. This has required to upgrade the DAQ systems for all the ALICE subdetectors, while for the Time Projection Chamber an additional major hardware upgrade has been implemented to allow for a continuous readout. 

Three new detectors have been installed. To achieve the needed vertexing resolution, the Inner Tracking System has been fully replaced, and a Muon Forward Tracker has been added in front of the Muon Spectrometer. Moreover, a new Fast Interaction Trigger detector system has been realized and put in place on both sides of the Interaction Point.

The large data flow arising from the new readout strategy is handled by the new Online-Offline (O$^2$) data processing, which reduces the amount of stored data by a factor 30 with respect to the input data flow.

ALICE is now working to further upgrade its  apparatus in view of the forthcoming Run~4, scheduled in 2029. The inner layers of the ITS  will be replaced with bent silicon detectors with very low material budget, and a high-granularity FOrward CALorimeter will enable a high-precision inclusive measurement of direct photons and jets.

% Authors with no connection to \LaTeX{} should use this
%sample text as a guide for their presentation using their favorite
%text editor 

%\subsection{Producing the Hard Copy}\label{subsec:prod}

\subsection{The Inner Tracking System 2}\label{subsec:ITS}

\begin{figure}[h]
\centerline{\includegraphics[width=0.7\linewidth]{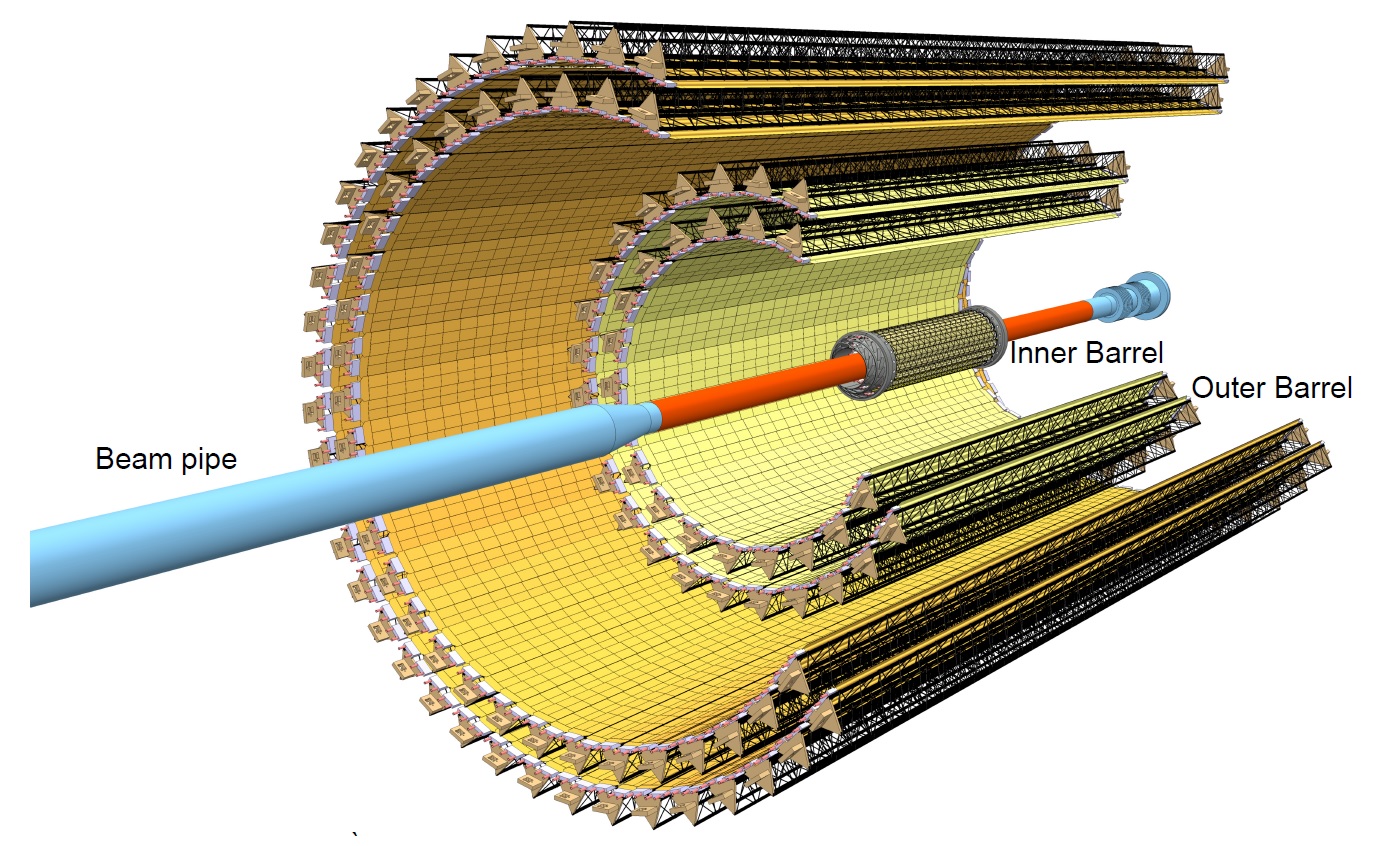}}
\caption[]{Layout of the ITS2 detector}
\label{fig:itsim}
\end{figure}

The ITS2\cite{its} is made of 7 cylindrical layers of silicon pixel detectors, arranged coaxially to the beam line and centered on the IP (fig.\ref{fig:itsim}). It has been designed to improve both vertex and tracking precision: to this end, the innermost layer radius is as small as 22 mm, almost halved with respect to the 39 mm in ITS1.  It is based on the ALPIDE chip (Monolithic Active Pixel Sensor, MAPS\cite{maps}), which performs in-pixel amplification, shaping, discrimination and multi-event buffering while absorbing  less than 40 mW/cm$^2$ of specific power. The pixel pitch of 27$\times$29 $\mu $m$^2$ and the detection efficiency is $>$99\%; it can withstand a particle rate of  $~100$ MHz/cm$^2 $ without pile-up.

ITS2 will provide high accuracy on secondary vertex determination due to  the extremely low material budget (below $0.35 \% X/X_0$ for the 3 innermost layers). Impact parameter resolution will be better by a factor of 3 and low $p_{\rm t}$ detection will start from 50 MeV/c; charm and beauty mesons will be measured  down to zero $p_{\rm t}$. Detector construction started in December 2016 involving more than ten production sites and 30 institutes worldwide, and ended with the installation of the inner barrel in May 2021.

\subsection{The Muon Forward Tracker}\label{subsec:MFT}

The MFT\cite{mft} is a new high-resolution silicon tracker covering the rapidity interval 2.5$<\eta<$3.6). It is placed before the front absorber of the Muon Spectrometer, and it will add precise vertexing capabilities to the Muon Tracking system thanks to the matching of muon tracks downstream of the absorber with MFT tracks. The impreoved resolution will allow to perform charm/beauty separation via secondary vertex reconstruction, and to provide a robust  $\psi$(2S) measurement by improving the signal to background ratio by a factor between 5 and 6.

It is based on the same ALPIDE chips as ITS2: it includes 936 chips (about 0.4~m$^2$ of detection area) arranged in 10 half-disks equipped with 2 detection planes each, for a total of 10 detection layers. Installation has been completed in December 2020.

\subsection{The Time Projection Chamber upgrade}\label{subsec:TPC}

The ALICE TPC\cite{tpc} is 5~m long, and its drift time lasts 100~$\mu$s. Up to the end of Run 2, the readout of the electrons generated inside the ALICE TPC active volume was performed by Multi-Wire Proportional Chambers: this limited the readout rate to few kHz, because of the need of a wire gating grid used to minimize ion backflow into the TPC. 
The goal of the upgrade was to achieve continuous readout, keeping in mind that a nominal gain =2000 was needed with the TPC Ne-CO$_2$-N$_2$ gas mixture, that the ion back-flow had to be kept under 1\%, and that the dE/dx performance had to be preserved.

The solution has been found in replacing the gated  MWPC with quadruple Gas Electron Multipliers stack: the external foils are standard hole pitch (140 $\mu$m), while the two internals are made with large hole pitch (280 $\mu$m) GEM foils.

After amplification in the stack, the signal charge is processed by newly developed  front-end ASIC SAMPA\cite{sampa}, made with 130 nm TSMC CMOS technology, each equipped with 32 channels, preamplifier, shaper and 10 bit ADC. Data are sent to the Front-End Cards (FECs), which receive the output of 5 SAMPA chips in continuous sampling mode at 5 MHz. There are 3276 FECs, and all the ADC values are read out. The data output for 50~kHz Pb-Pb collisions amounts to 3.3~TByte/s: from this value, it is clear the need of data filtering/compressing before storage (see section~\ref{subsec:O2}). 

The construction started in August 2016 with GEM production. In 2019 TPC was brought on the surface: installation of GEM modules started in May, pre-commissioning in November and final installation inside L3 magnet ended in December 2020.

\subsection{The Fast Interaction Trigger Detector}\label{subsec:FIT}

The new FIT\cite{fit} is a composite detector system placed along the beam line that is essential, in many ways, for the operation of ALICE. It is made by three sub-systems: FV0 is a scintillator-based disk-shaped detector read out by optical fibers. It is placed 3.5 m from the IP on the opposite side of the Muon Spectrometer, covering the rapidity interval 2.2$<\eta<$5.0. It measures multiplicity, centrality and event plane of the ion collisions, and contributes to the minimum bias and centrality triggers with latency $<$425~ns that are used for online vertex determination, selection and rejection of beam-gas events. 

FT0-A and FT0-C are two quartz Cherenkov radiators arranged in squares of side of about 20 cm, read out by microchannel plate-base photomultipliers and placed on opposite sides with respect to the IP. FT0-A is placed immediately behind FV0 and FT0-C is 80 cm from the IP, between the MFT and the muon absorber. The FT0 main task is to measure the collision time that is then used in Time Of Flight-based particle identification and, together with FV0, to issue the above-mentioned selection triggers. 

Finally, FDD-A and FDD-C are placed on both sides of the IP, respectively at 17~m and 19.5~m distance, with rapidity acceptance extending up to $\eta$=6.9. They are made of plastic scintillator pads coupled with wavelength shifting bars and optic fibers for readout. They provide a tag for diffractive events, measurements of the diffractive cross-sections and of ultra-peripheral collisions products. 
The installation of the FIT in the ALICE cavern was finalized in June 2021.

\subsection{Online-Offline processing: O$^2$ }\label{subsec:O2}

The O$^2$ processing\cite{o2} has been devised to cope with the 3~TByte/s flow of data, reducing it to a storage output of 0.1 TByte/s while preserving physics and monitoring information.
  
It is made of First Level Processors (FLPs) which receive detector data from detectors and chop the
continuous data flow into (sub-)time frames lasting 10-20 ms each. These are sent to the Event Processing Nodes, which filter the flow down to a manageable 0.1~TByte/s that are sent to disk storage. 
All the process is regulated by the Central Trigger Processor, which provides the common clock, the syncronization signals (including triggers to detectors non implementing continuos readout) and the timing information needed for the correct assembly of the sub-events.

\subsection{ITS3 for Run 4 }\label{subsec:ITS3}

The performance of the ITS2 can be further improved essentially by bringing its Inner Barrel closer to the Interaction Point. This can be done by installing a beam pipe with smaller inner radius (from 18.2 mm to 16 mm) and reduced thickness (from 800~$\mu$m to 500~$\mu$m).
Another way to improve drastically the ITS physics performance is to lower its material budget. To this end, ALICE is testing 65~nm CMOS silicon detectors, thinned down up to the point that they become flexible (20-40 $\mu$m) and stitched together to create large sensors up to 300 mm. 

Thanks to a power consumption below 20 mW/cm$^2$ due to the 65 nm technology, the water cooling system is no longer necessary. The integration of power and data buses directly on sensor chips allows discarding the circuit board, and if these low-weight sensors are bent to target radii, they acquire sufficient stiffness to remove the need of a complex support structure: they can be mechanically held in place by low-density carbon foam ribs.
The main benefits of these improvements are a very low material budget (from 0.34\%$~X/X_0$ to 0.04\%~$X/X_0$) and a very homogeneous material distribution which reduces the systematic error to a negligible level. A pixel size of the order of $10 \times 10$ $\mu$m$^2$ will lead to further resolution improvement.

\subsection{FOrward CALorimeter for Run 4}\label{subsec:FOCAL}

The ALICE collaboration is working towards the realization of a a new, high-granularity Forward Calorimeter able to perform high-precision inclusive measurements of direct photons and jets, as well as coincident $\gamma$-jet and jet-jet measurements. It will be placed along the beam line, 7~m from the Interaction Point, covering the rapidity interval 3.4$< \eta < 5.8$. It will be made of two sub-detectors: FoCal-E is a high-granularity (1~mm$^2$) Si-W sampling sandwich electromagnetic calorimeter for photons and $\pi^0$, equipped with pads and 2 high-granularity pixel layers, while FoCal-H is a conventional sampling hadronic calorimeter made of copper and scintillating fibres, for photon and jets isolation. Installation is foreseen in time for the start of Run 4 in 2029.

\section{Conclusions}

In view of the imminent start of Run 3, all the major ALICE upgrades for Run 3 are on track. They consist in a full upgrade of the detector readout architecture and computing and in a new integrated Online-Offline system. Three new detectors have been installed: ITS2 and MFT will provide enhanced tracking and vertexing performance, while the new, GEM-based TPC Readout Chambers will allow a 50-fold increase of the event readout rate. A new Fast Interaction Trigger will help in the process of data selection. In view of Run 4, two new upgrades will allow further improvements of the the physics performances of the ALICE detectors from 2029 onwards.

\end{document}